\documentclass[prl,superscriptaddress,twocolumn,showpacs]{revtex4}
\usepackage{epsfig}
\usepackage{amsbsy}
\usepackage{amsmath}
\usepackage{amsfonts}
\usepackage{graphics}
\usepackage{thumbpdf}
\usepackage{latexsym}
\usepackage{bm}

\begin{document}

\title{Entanglement dynamics in chaotic systems}  
\author{Shohini Ghose}
\affiliation{Institute for Quantum Information Science, University of Calgary, Alberta T2N 1N4,
Canada}
\affiliation{Department of Physics and Astronomy, University of New Mexico, Albuquerque, NM 87131, USA}
\author{Barry C. Sanders}
\affiliation{Institute for Quantum Information Science, University of Calgary, Alberta T2N 1N4,
Canada}\
\affiliation{Centre for Quantum Computer Technology,
Macquarie University, Sydney, New South Wales, Australia}

\date{\today}
\begin{abstract}
We study quantum chaos for systems with more than one degree of
freedom, for which we present an analysis of the dynamics of entanglement.
Our analysis explains the main features of entanglement dynamics and identifies
entanglement-based signatures of quantum chaos. We discuss entanglement
dynamics for a feasible experiment involving an atom in a magneto-optical
trap and compare the results with entanglement dynamics for the
well-studied quantum kicked top.
\end{abstract}
\pacs{05.45.Mt, 03.65.Ud, 03.67.-a, 39.25.+k}
\maketitle

Theories of chaos and of quantum mechanics juxtapose in the discipline
of `quantum chaos' (QC),
which has attracted significant theoretical and experimental~\cite{QChaos, Exp, Exp2}
efforts. Recently the advent of quantum information~\cite{Nie00} has highlighted the role of entanglement as a resource, and stimulated theoretical studies of entanglement in QC systems~\cite{entropy1,entropy2,entropy3,WGSH2003,L2003, J2003}. 
Here we present a general analysis of entanglement dynamics for unitarily evolving QC systems, 
and apply this analysis to a system of significant experimental interest:
an atom in a magneto-optical lattice (AMOL), which we show is a feasible
experimental system to observe and test entanglement dynamics and to rapidly enhance entanglement for certain initial states. The theoretical methods and results are general, which we demonstrate by application to the well-studied quantum kicked top (QKT)~\cite{WGSH2003,QKT,QKT1,San89}. We also identify the initial entanglement rate as a signature of quantum chaos.
 
Entanglement features of QC systems are rather subtle:
in fact recent theoretical studies reveal a dichotomy: entanglement may be enhanced as the chaoticity parameter is increased~\cite{entropy1,WGSH2003} or the opposite can occur~\cite{entropy3}. In fact, this dichotomy can be understood by observing that the former result applies for a global chaotic quantum system for which entanglement applies between its constituents, whereas in the latter case one considers weak coupling between a quantum chaotic system and the environment which could be another quantum chaotic system. Our focus is on the former case which can yield a rapid increase of entanglement.

In particular, we analyze the
evolution of an AMOL~\cite{MOL} under
realistic experimental conditions and show that entanglement arises between the atomic spin and motional degrees
of freedom. The ability to tomographically reconstruct the reduced density matrix of the atomic spin state~\cite{QST} makes it possible to study the evolution of this entanglement experimentally. We analyze the dynamics of entanglement (via the linear entropy of the spin density matrix) and characterize the global dynamics by the size of the chaoticity
parameter and local dynamics by whether the initial state is supported primarily by
regular or chaotic eigenstates of the Hamiltonian. 
AMOL experiments would allow the first empirical studies of
entanglement evolution in a QC system. 

An AMOL provides an attractive framework for studying entanglement evolution in a QC system with and without coupling to an environment. This is because, in addition to having more than one degree of freedom and the ability  to tomographically reconstruct states, decoherence can be controlled by detuning the laser relative to the atomic resonance frequency. 
Decoherence is negligible for the far off-resonance AMOL
so coupling to the environment can be ignored (unitary dynamics), and entanglement between  
the spin and motional subsystems of the overall chaotic system can be explored. By tuning the laser frequency close to atomic resonance, coupling to the environment is increased and the resulting effects of entanglement can be observed.

Here we are concerned with the former case of the far off-resonance magneto-optical lattice for which coupling to the environment can be neglected and  entanglement is enhanced. We show that under feasible conditions, the AMOL can exhibit generic features of entanglement dynamics, for example quasiperiodicity for a state initially localized in a regular regime and a rapid increase of entanglement with no subsequent quasiperiodicity in a chaotic regime. 

Generic features of entanglement in QC systems can be understood by examining the spectral properties of the evolution operator $U(t)$ on the system Hilbert space $\mathcal{H} = \otimes_{i=1}^{N} \mathcal{H}^{(i)} $ with $\mathcal{H}^{(i)}$ the Hilbert space of dimension $d_i$ for the $i^{\text{th}}$ subsystem. We consider two common categories of unitary evolution:
(i)~$U(t)=\text{exp}(-iHt/\hbar)$ for a time-independent Hamiltonian $H$ and 
(ii)~$U(t=n\tau) = F^n$ with $F$ being a Floquet operator $F=T\text{exp}[-i/\hbar\int_0^\tau H(t) \text{d}t]$ 
and $t$ a discrete time variable. 
The unitary evolution operator can be expressed as $U(t)=\sum_i ^d\text{exp}(-\text{i}\omega_i t)|\phi_i\rangle\langle\phi_i|$ for  $\{|\phi_i\rangle\}$ a time-independent orthonormal basis of $\mathcal{H}$ of dimension $d$ and $\{\text{exp}(-\text{i}\omega_i t)\}$ the corresponding eigenvalues. The evolution of an arbitrary initial density operator $\rho(0)$ over time $t$ is
$U(t)\rho(0)U^\dagger(t) = \sum_{i,j} e^{-\text{i}\omega_{ij}t}\rho_{ij}|\phi_i\rangle\langle\phi_j|$
for $\omega_{ij}\equiv \omega_i-\omega_j$ and $\rho_{ij}\equiv \langle\phi_i|\rho(0)|\phi_j\rangle$. 

Entanglement for pure states ($\rho=\rho^2$) with two subsystems is given by the entropy of the reduced density operator $\tilde \rho(t)$ of either subsystem. The linear entropy $S=1-\text{Tr}(\tilde \rho^2)$ is a convenient measure of entanglement,
with $S=0$ for no entanglement and $S=1-1/d_i$ for maximum entanglement. The time-dependent entropy is 
\begin{equation}
\label{St}
S(t)= 1-\sum_{i,j,k,l}
C_{ijkl}\text{e}^{-\text{i}(\omega_{ij}+\omega_{kl})t}
\end{equation}
with
\begin{eqnarray}
C_{ijkl}=\rho_{ij}\rho_{kl} \sum_{m,n,p,q} & &\langle u_p,v_m|\phi_i\rangle\langle\phi_j|u_p,v_n\rangle\nonumber\\
& &\langle u_q,v_n|\phi_k\rangle\langle\phi_l|u_q,v_m\rangle
\end{eqnarray}
for $\{|u_k\rangle\},\{ |v_m\rangle\}$ orthonormal bases for $\mathcal{H}^{(1)}$ and $\mathcal{H}^{(2)}$.

Diagonal elements $\rho_{ii}$ quantify support of $\rho$ on $U$-eigenstates $|\phi_i\rangle$. These eigenstates can be associated with regular and chaotic regimes~\cite{P1994, E1917};
hence a state $\rho$ can be identified with classical regular or chaotic regimes (or a combination)
by its support on $U$-eigenstates.
We employ this correspondence between support~\cite{QKT1} and regular vs chaotic dynamics 
to characterize entanglement dynamics for an AMOL and other QC systems.   

The AMOL is  subjected to a far off-resonance AC Stark
shift as a function
of atomic position $z$ given by~\cite{MOL}
\begin{equation}\label{ACStark}
V = \frac{4}{3}V_1\cos \Theta _\text{L}\cos {2kz}
- \bm{\mu }\cdot \bm{B}_{\text{eff}}\left( z \right),
\end{equation}
for
$\bm{\mu}=\hbar\gamma\bm{F}={-\mu_\text{B}\bm{F}}/{F}$,
$\gamma$ the gyromagnetic ratio and $\bm{F}$ the total
angular momentum vector of the hyperfine ground state. As an example, we consider
$^{133}$Cs with $F=4$ and
$\mu _\text{B}\bm{B}_{\text{eff}}(z)=\frac{2}{3}V_1\sin \Theta _\text{L}\sin
2kz\,\bm{e}_z+B_x\bm{e}_x$ for $k$ the laser wave vector, $\Theta_L$ the relative polarization angle between the counter-propagating laser beams forming the 1-D lattice and $V_1$ the single-beam light shift. The coupling between the spin precession and center-of-mass motion leads to
entangled spinor wave packets and $B_x$ is the tunable chaoticity parameter. (In fact, Eq.~(4) is applicable to more general systems if the periodic potential is replaced by a harmonic potential~\cite{spinboson}.)
In the classical limit, Eq.~(3) describes a magnetic moment
interacting with the same $\bm{B}_{\text{eff}}(z)$~\cite{GAD2001}, with 
$\bm{F}/F$ the direction vector for the classical angular momentum, and the classical 
four-dimensional phase space parametrized by
atomic position and momentum $(z,p)$ and direction $(\theta,\phi)$, of $\bm{F}/F$.

We seek to study entanglement dynamics for lattice  parameters that are accessible in current
experiments. Therefore we choose $V_1=160E_R$, $\Theta_L=80^{\circ}$ and
$\mu_BB_x=12E_R$, for $E_R=\hbar^2k^2/2M$, the recoil energy. 
Classical Poincar\'{e} sections for these parameters and total
energy  $E= p^2/2M + V =-280E_R$ reveal a mixed phase space with islands of regular motion embedded in
the chaotic sea (Fig.~1). Quantum states are localized
to phase space coordinates $(z,p,\theta,\phi)$ by preparation in
a product of the motional and spin coherent states
$|\alpha=z+ip\rangle|\theta,\phi\rangle$~\cite{SCSS}. 
 \begin{figure}
\includegraphics[width=0.45\textwidth]{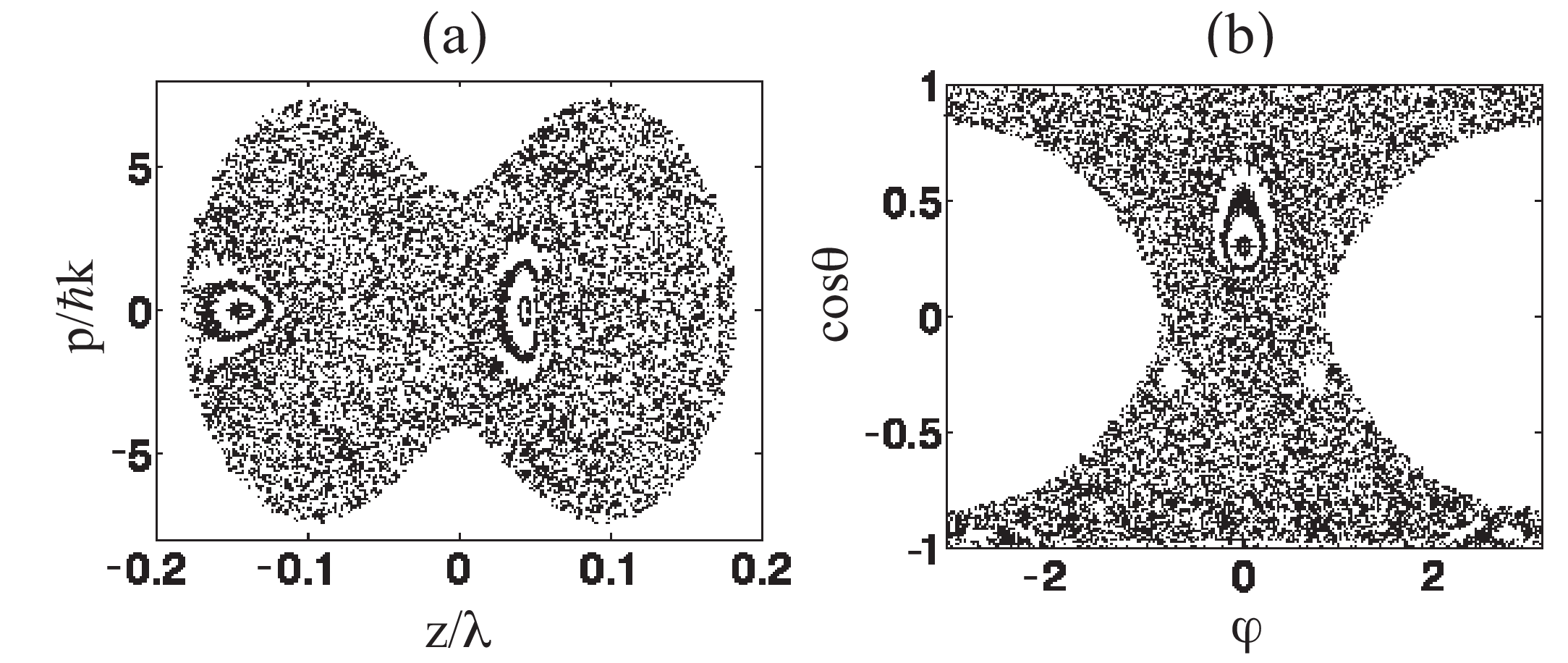}
\caption{Classical Poincar\'{e} sections for the AMOL at $E=-280E_R$ for
$V_1=160E_R$, $\Theta_L=80^o $ and $\mu_BB_x=12E_R$ with 
 (a)~$\mu_y=0, \text{d}\mu_y/\text{d}t >0$ and 
 (b)~$p=0, \text{d}p/\text{d}t >0$.}
\end{figure}

An AMOL state localized around $(z,0,\theta, \phi)$ can be prepared by cooling atoms to the ground
state of the diabatic potentials. The lattice is then shifted
until this state is centered at   $(z/\lambda)$. The spin is rotated until the Bloch vector is pointing in the direction
$(\theta,\phi)$. We pick an initial state that is centered on an island in the Poincar\'{e} section  in Fig.~1. For comparison we also consider an initial state in the chaotic sea. 
\begin{figure}
\includegraphics[width=0.45\textwidth]{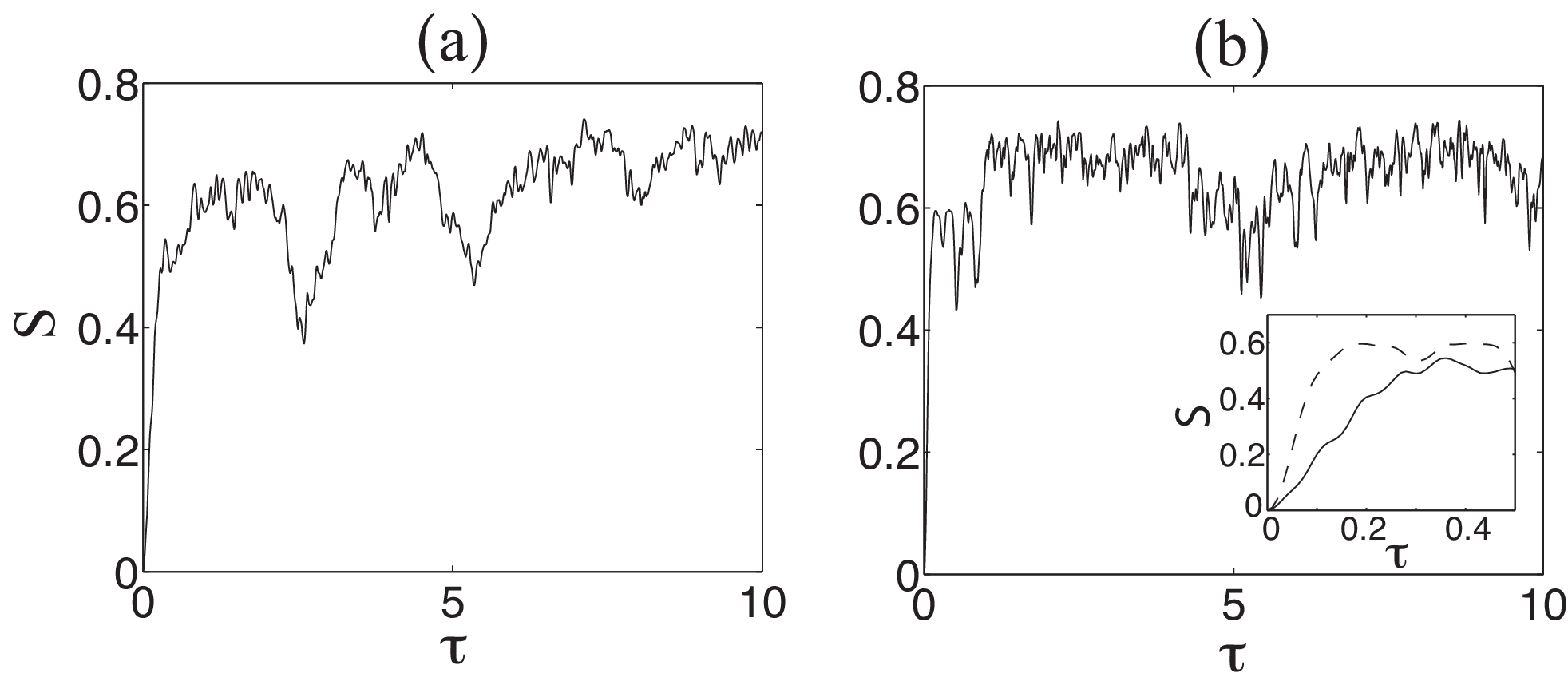}
\caption{Entanglement $S$ versus time $\tau=E_Rt/\hbar$ in the AMOL for an
initial state that is localized on (a) a regular island with $(z/\lambda,p/\hbar k, \theta, \phi) = (-0.15,0,1.27,0)$ and (b)
the chaotic sea with  $(z/\lambda,p/\hbar k,\theta,\phi)=(0.06,0,\pi/2,0)$. The inset shows the initial increase of entanglement for the regular (solid) and chaotic (dashed) initial states.}
\end{figure}

The evolution of entanglement, quantified by linear entropy, for
states that are initially regular (Fig.~2(a)) or chaotic (Fig.~2(b)), exhibit two main
signatures of chaos.
At short times, entanglement in the chaotic regime increases at a faster rate than
for the regular regime, thereby 
supporting the concept that chaos can cause rapid generation of
entanglement as observed in other systems~\cite{entropy1,WGSH2003}.
Also oscillations are prevalent for initially regular states but not for chaotic states
(which has been also observed for the quantum kicked
top~\cite{WGSH2003}).  We explain how the power spectrum of $S$ can provide
a signature of chaos, as Lahiri suggested~\cite{L2003}, by exploiting the
$U$-eigenbasis.
\begin{figure}
\includegraphics[width=0.45\textwidth]{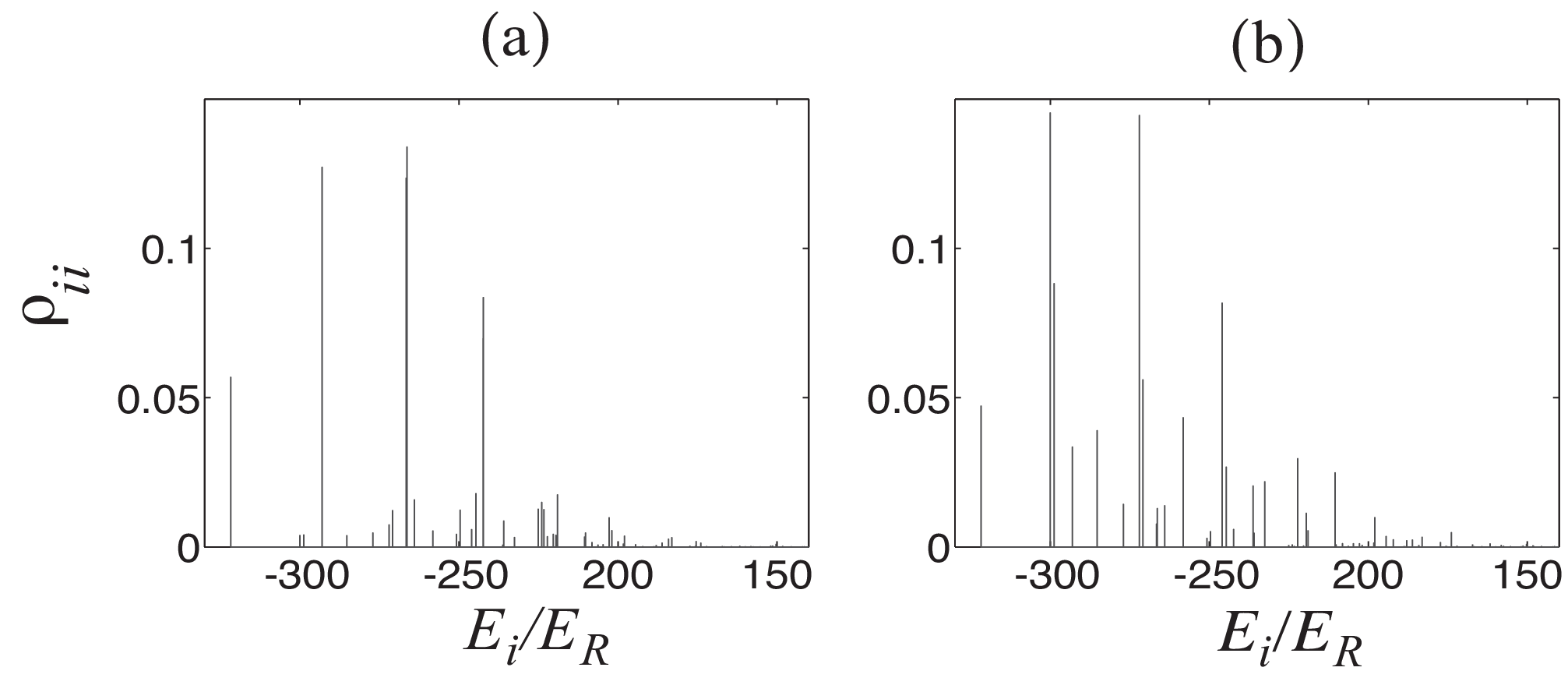}
\caption{Population $\rho_{ii}$ vs corresponding eigenenergy $E_i=\hbar\omega_i$ for the AMOL with an initial state localized in (a) a regular island with $(z/\lambda,p/\hbar k, \theta, \phi) = (-0.15,0,1.27,0)$ and (b)
the chaotic sea with  $(z/\lambda,p/\hbar k,\theta,\phi)=(0.06,0,\pi/2,0)$.}
\end{figure}

$U$-eigenstate support for the states initially in the regular and chaotic regimes are depicted in
Fig.~3. The initial state on the regular island has support dominated by four pairs of regular eigenstates
with each pair nearly degenerate. This support over few eigenstates is responsible for 
the quasiperiodic evolution of linear entropy, and Fig. 4(a) shows that excellent replication of
entanglement dynamics is possible by only including these four pairs of eigenstates.
In contrast, the initial state in the chaotic sea
has support over a larger number of the chaotic set of eigenstates extending over a broader frequency spectrum (Fig.~3(b)), due to a breakdown of semiclassical theory in the chaotic regime~\cite{P1994, E1917}. 

A rigorous understanding of entanglement evolution emerges by noting that
$S(t)$ in Eq.~(\ref{St}) depends on eigenfrequency-difference sums 
$\omega_{ij}+\omega_{kl}$, which can be identified in the power spectrum of $S(t)$. The fast oscillations in Fig.~2(a) are due
to the large differences $\delta \omega$ between the four main
peaks in Fig.~3(a), and small frequency differences between
almost degenerate eigenstate pairs at each peak in Fig.~3(a) result in
slow oscillations with long periods as seen in the long-term behavior of the entanglement in Fig.~4(b). The terms~$\omega_{ij}+\omega_{kl}$ that appear in the evolution correspond not just to differences in the eigenfrequencies, but also can be a sum of $\omega_{ij}+\omega_{kl}$.  For example, the main oscillation in Fig.~2(a) is due to the sum of two difference frequencies.
\begin{figure}
\includegraphics[width=0.45\textwidth]{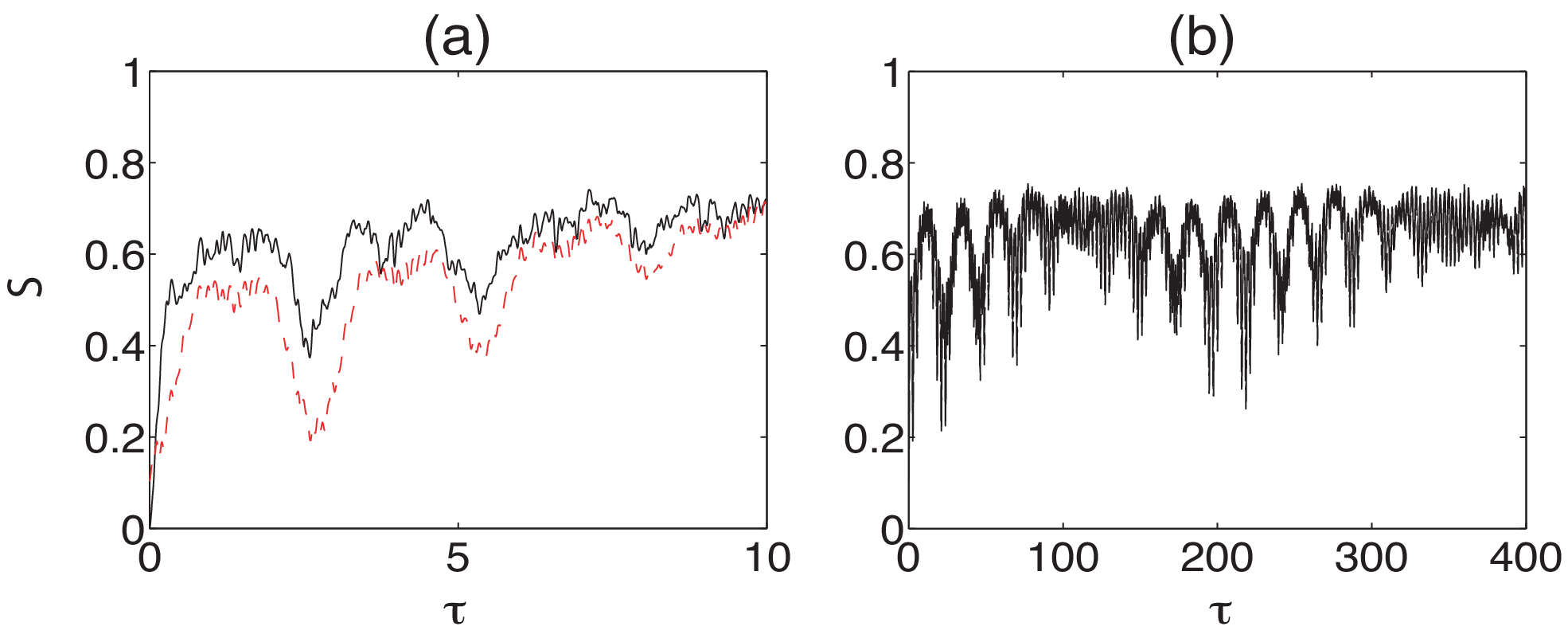}
\caption{(a) For the initial state localized on the regular island the entanglement dynamics (solid curve) can be reproduced by only considering the evolution of the four main pairs of eigenstates (dashed curve). The long term behavior (b) shows quasiperiodic motion with multiple frequencies.}
\end{figure}

The key feature of dynamical entanglement for our purpose is
the  initial increase of entanglement for a chaotic state,
which is more rapid than for  the regular state (inset of Fig.~2).
This rapid rise in entanglement is not unique to the AMOL system~\cite{entropy1, WGSH2003}.
The rate of increase of entanglement is obtained by expanding Eq.~(1) at $t=0$,
which reveals a quadratic increase as a function of time, $S=(t/t_0)^2$ with $t_0=0.01$; this behavior is
surprising at first because an exponential increase is expected for states in the chaotic regime
and a quadratic increase for states in the regular regime~\cite{J2003}.
Of course this expectation applies for the asymptotic semiclassical regime, but our system lies in a deeply quantum regime and hence the
chaotic state is not well localized, with non-negligible support over the regular regime which results in the
quadratic increase in entanglement. For the initial state in the chaotic regime, $\Delta x/\lambda \approx 0.07$, $\Delta p/\hbar k \approx 2.7$,  $\Delta \mu_x=0$ and $\Delta \mu_z=\Delta \mu_y = 1/\sqrt 2$. 

Our methods to analyze the AMOL are generic and generalize to other unitarily evolving QC systems as we now show for the quantum kicked top with Hamiltonian~\cite{QKT,QKT1,San89}
\begin{equation}
	H=\frac{\kappa}{2j\tau}J_z^2+pJ_y\sum_{n=-\infty }^{\infty }\delta
(t-n\tau ),  \label{top}
\end{equation}
for $J_x,J_y,J_z$ su(2) operators and $\kappa$ the chaoticity parameter. 
The QKT  can be constructed from a collection of $N=2j$ qubits in the symmetric representation with collective spin operators $J_{\alpha }=\sum_{i=1}^{N}\frac{\sigma _{i\alpha }}{2}$, and $\{\sigma_{i\alpha}\}$ the Pauli operators for the $i^{\text{th}}$ qubit~\cite{WGSH2003}.  For $\kappa=3, \tau=1, p=\pi/2$ entanglement behavior is similar to that of the trapped atoms described here~\cite{WGSH2003}.  Bipartite entanglement between a pair of qubits and the remaining qubits reveals quasiperiodic evolution for an initial state
centered on an elliptic fixed point. For a state centered in the
chaotic sea, no quasiperiodic motion is present, and just as in our AMOL, an initial rapid increase of entanglement is observed confirming this generic behavior.

Support of an initial state over $(U=F^n)$-eigenstates,
of the Floquet operator $F=\text{exp}(-\text{i}\kappa J_z^2/2j\tau)\text{exp}(-\text{i}pJ_y)$
is shown for an initial state centered
on the elliptic fixed point and one in the
chaotic region of Ref.~\cite{WGSH2003}. The
initial state centered in the regular region can be mainly decomposed
into a few (size $\sqrt{N}$)  `regular' eigenstates~\cite{QKT1}. In this case the state which we have localized at a fixed point has most of its support on three eigenstates, of which two are degenerate. The difference between the corresponding eigenphases $\phi_m$, $\delta\phi=0.003$ determines the frequencies of oscillation in the
evolution of $S$ (Fig.~6(a)).
\begin{figure}
\includegraphics[width=0.45\textwidth]{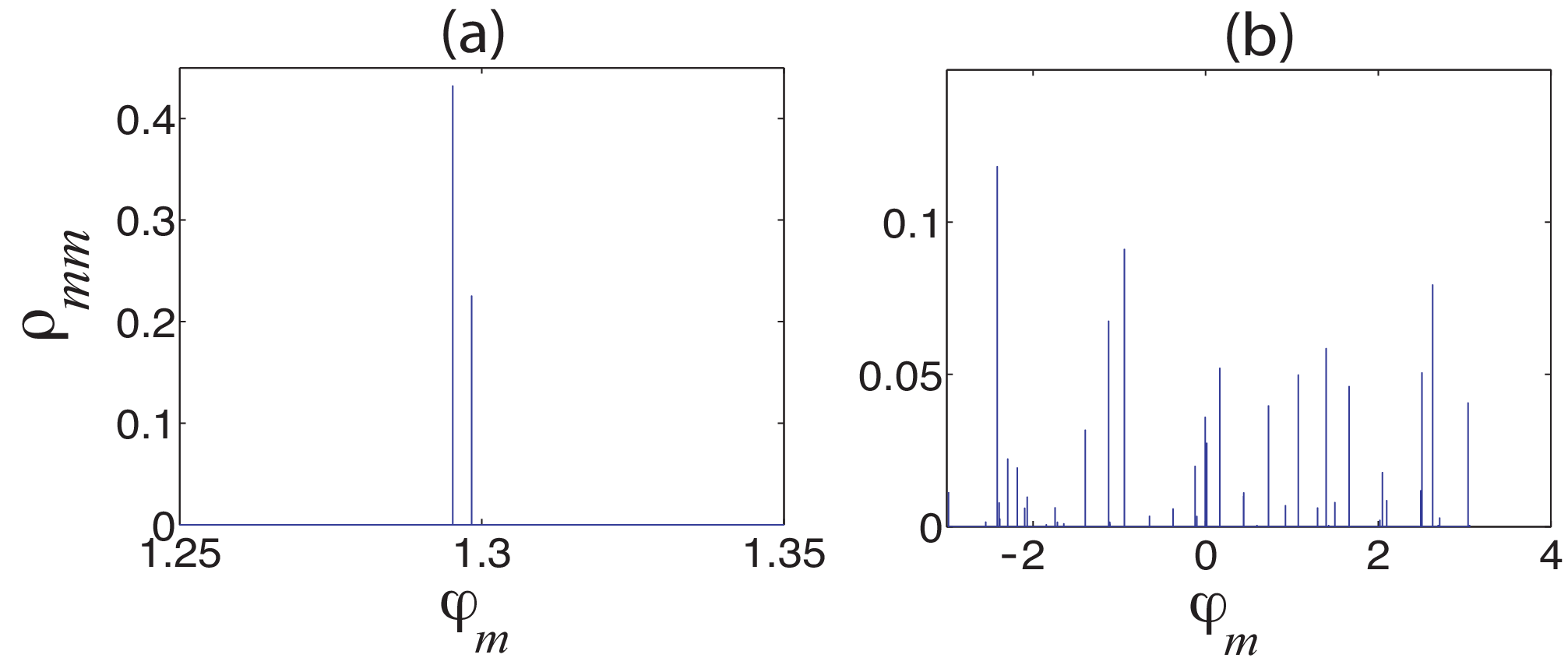}
\caption{Population $\rho_{mm}$ vs corresponding eigenphases $\phi_m$ for the QKT with an initial state localized in (a) a regular island and (b)
the chaotic sea in Fig. 1 of~\cite{WGSH2003}.}
\end{figure}
  
The flat entanglement power spectrum for the chaotic state is due to the broad
support (size $N$)~\cite{QKT1} of the initial state on the `chaotic'  $U$-eigenstates
(Fig.~5(b)). Because $N=50$ qubits is in the semiclassical regime, the distinction between
regular and chaotic entanglement dynamics is more pronounced than
what we observed for the AMOL, which was not as semiclassical. (As we consider $^{133}\text{Cs}$ for the AMOL, $F=4$ is fixed and thus not in the semiclassical regime).
Moreover the rise time for the initial chaotic state is exponential~\cite{J2003} which can be regarded as a signature of quantum chaos.
\begin{figure}
\includegraphics[width=0.45\textwidth]{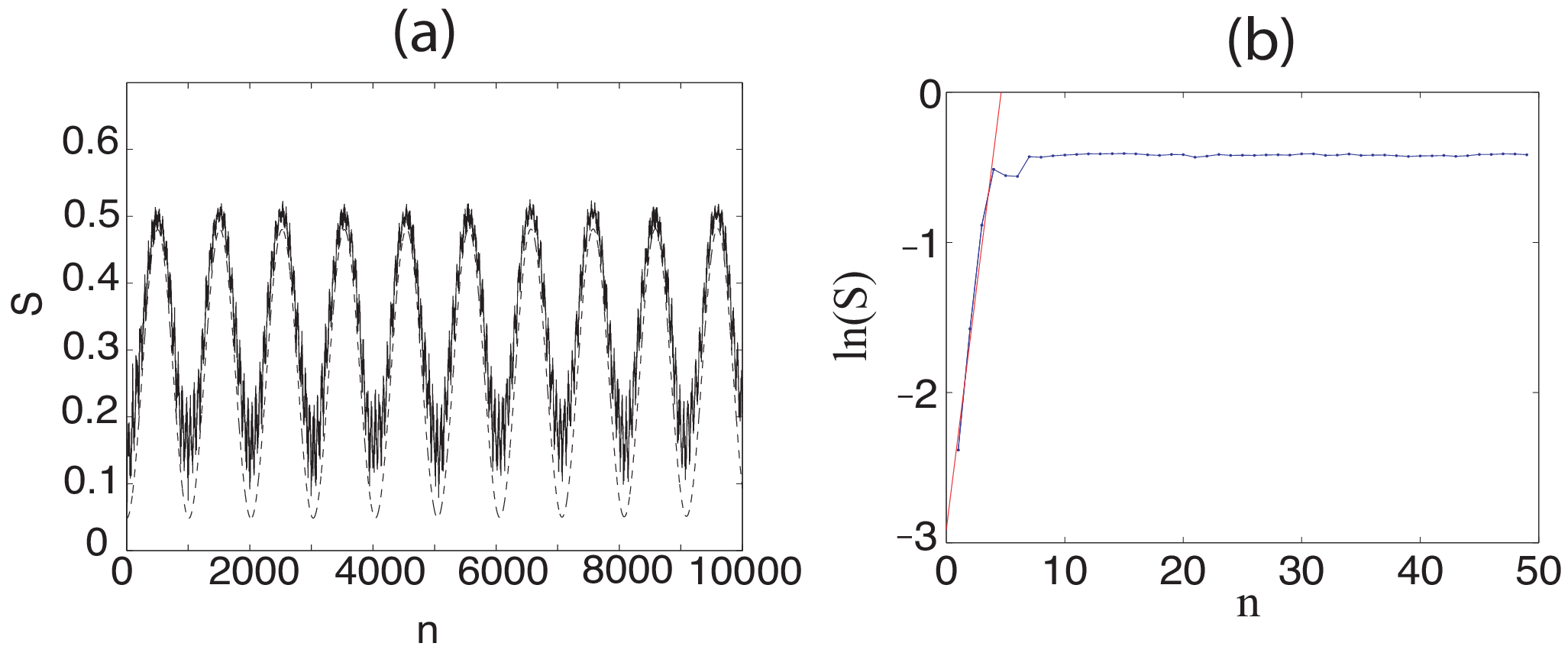}
\caption{(a)~For the regular initial state of the QKT the entanglement dynamics (solid curve) can be reproduced by considering the evolution of the three highest weighted eigenstates (dashed curve) in Fig. 5(a). (b)~The rise time for the state in the chaotic sea is roughly exponential.}
\end{figure}

The QKT system of $N$ qubits behaves collectively like a system with one degree of freedom unlike the AMOL which has two degrees of freedom: spin and motion. Coupling between two QKTs, however,
allows for entanglement dynamics between two coupled degrees of freedom to be observed,
but entanglement is suppressed in the strong chaotic regime~\cite{entropy3}.
This reduction of entanglement  can be explained by noting that
our results imply that the entanglement between qubits in one QKT is chaos-enhanced,
which causes the qubits to be collectively less entangled with the qubits of the other QKT.

In conclusion, we have presented entanglement dynamics for an experimentally feasible QC system of atoms trapped in a magneto-optical lattice. For realistic experimental parameters, quantum signatures of chaos exist in the dynamics of entanglement, specifically in the initial rise and the power spectrum, even when the system is not in a semi-classical regime, and there is a rapid increase of entanglement for initial states in the chaotic 
regime. Our analysis relies on studying support over the unitary evolution eigenbasis, which
applies generically to other unitarily evolving QC systems, as we demonstrate 
with the quantum kicked top. Thus we have introduced a means for extending QC
experiments to more than one degree of freedom, observing and understanding entanglement
dynamics in such systems, exploiting quantum chaos for a rapid increase in entanglement,
and resolving the dichotomy between cases where chaos enhances vs 
diminishes entanglement generation for initially chaotic states.

\acknowledgements
We appreciate valuable discussions with I. Deutsch, P. Jacquod and Xiaoguang Wang (who also provided computer codes for obtaining data in Fig.~6). Numerical results for the AMOL system were partially based on codes developed by J. Grondalski. This project has been supported by an Australian Research Council Large Grant and Alberta's Informatics Circle of Research Excellence (iCORE) and the National Science Foundation under  Grant No. PHY-009569.

\end{document}